\documentclass[sigconf]{acmart}

\usepackage{url}

\usepackage{subcaption}
\usepackage[capitalise]{cleveref}
\usepackage{tabularx}
\newcolumntype{L}{>{\raggedright\arraybackslash}X}
\newcolumntype{R}{>{\raggedleft\arraybackslash}X}

\copyrightyear{2026}
\acmYear{2026}
\setcopyright{cc}
\setcctype{by}
\acmConference[ICPE Companion '26]{Companion of the 17th ACM/SPEC International Conference on Performance Engineering}{May 04--08, 2026}{Florence, Italy}
\acmBooktitle{Companion of the 17th ACM/SPEC International Conference on Performance Engineering (ICPE Companion '26), May 04--08, 2026, Florence, Italy}
\acmDOI{10.1145/3777911.3800636}
\acmISBN{979-8-4007-2326-1/2026/05}

\begin{document}

\title{Performance Optimization in Stream Processing Systems: Experiment-Driven Configuration Tuning for Kafka Streams}

\author{David Chen}
\orcid{0009-0009-1386-9083}
\affiliation{%
  \institution{Johannes Kepler University} %
  \city{Linz}
  \country{Austria}
}
\email{chendavidyiwen@gmail.com}

\author{Sören Henning}
\orcid{0000-0001-6912-2549}
\affiliation{%
  \institution{Dynatrace Research}%
  \city{Linz}%
  \country{Austria}%
}
\email{soeren.henning@dynatrace.com}

\author{Kassiano Matteussi}
\orcid{0000-0002-9131-6849}
\affiliation{%
    \institution{JKU/Dynatrace Co-Innovation Lab,\\Johannes Kepler University Linz}%
    \city{Linz}
    \country{Austria}
}
\email{kassiano.matteussi@jku.at}

\author{Rick Rabiser}
\orcid{0000-0003-3862-1112}
\affiliation{%
    \institution{LIT CPS Lab,\\Johannes Kepler University Linz}%
    \city{Linz}%
    \country{Austria}%
}
\email{rick.rabiser@jku.at}

\begin{abstract}
Configuring stream processing systems for efficient performance, especially in cloud-native deployments, is a challenging and largely manual task. We present an experiment-driven approach for automated configuration optimization that combines three phases: Latin Hypercube Sampling for initial exploration, Simulated Annealing for guided stochastic search, and Hill Climbing for local refinement. The workflow is integrated with the cloud-native Theodolite benchmarking framework, enabling automated experiment orchestration on Kubernetes and early termination of underperforming configurations. In an experimental evaluation with Kafka Streams and a Kubernetes-based cloud testbed, our approach identifies configurations that improve throughput by up to 23\% over the default. The results indicate that Latin Hypercube Sampling with early termination and Simulated Annealing are particularly effective in navigating the configuration space, whereas additional fine-tuning via Hill Climbing yields limited benefits.
\end{abstract}



\maketitle

\section{Introduction}

Stream processing systems play a crucial role in enabling near-real-time data analytics across various domains such as finance, e-commerce, IoT, and software monitoring~\cite{Fragkoulis2023}. Such systems continuously filter, transform, and aggregate data streams at high throughput and with low, often sub-second, latency~\cite{SEAA2023}. Driven by ever-growing data volumes and the availability of scalable compute resources in the cloud, stream processing systems are often implemented as distributed systems, requiring critical properties such as scalability, fault tolerance, resource efficiency, state management, and data partitioning to be addressed.

Open-source stream processing frameworks such as Apache Kafka Streams~\cite{Wang2021}, Apache Flink~\cite{Carbone2015}, and Apache Spark~\cite{Armbrust2018} provide built-in support for such requirements.
However, these frameworks expose numerous configuration parameters related to, for example, memory management, buffering, batching, or caching, which can substantially affect performance attributes such as  latency, throughput, scalability, and reliability~\cite{Fischer2015,Jamshidi2016,Bilal2017,vanDongen2020,DeSouza2020,BDR2021,Matteussi2022,Wang2025}. Deploying and operating such systems for high performance is therefore challenging,
especially in containerized cloud environments.
Given the sheer number of parameters and their intricate interdependencies, exhaustively exploring all configuration combinations is infeasible in practice. As a result, practitioners often rely on default settings, expensive expert knowledge, or ad hoc trial-and-error tuning, leaving considerable performance potential undiscovered.

In this paper, we propose an experiment-driven approach that uses heuristic methods from search-based software engineering to explore the vast configuration space of stream processing systems and identify performance-improving configurations.
We evaluate our approach with Kafka Streams, one of the most popular stream processing frameworks.
Using the ShuffleBench~\cite{ICPE2024} benchmark, we assess 265~different configurations in a Kubernetes-based deployment on a cloud testbed and observe throughput improvements of up to 23\% over the default configuration.

Our contributions are threefold: (i) an automated optimization approach that combines Latin Hypercube Sampling, Simulated Annealing, and Hill Climbing, (ii) a cloud-native integration with the Theodolite~\cite{EMSE2022} benchmarking framework for automated experiment orchestration on Kubernetes, and (iii) an experimental pilot evaluation that provides insights into the effectiveness of the individual optimization components.

\section{Background and Related Work}\label{sec:related-work}

Related work on stream processing performance evaluation has shown that framework configuration can significantly impact performance. For example, memory-related parameters affect throughput in Spark~\cite{Matteussi2022} and the mapping of logical components to physical resources can introduce contention, skew, and slowdowns in Flink~\cite{Wang2025}, or even latency between data sources and cloud providers \cite{Kyrama2026}. %
Building on these insights, we first discuss key approaches from the literature on configuration optimization for stream processing systems, before introducing three optimization techniques employed in this work.

\subsection{Parameter Optimization Approaches}

\citet{Herodotou2020} classified automatic parameter optimization approaches for big data processing systems, with the distinction between experiment-driven and adaptive approaches being particularly relevant for this work.

\subsubsection*{Experiment-driven Approaches}

Experiment-driven approaches rely on repeated executions of applications with different parameter settings, guided by search-based algorithms and results from actual runs.
For example, \citet{Fischer2015} and \citet{Jamshidi2016} investigated Bayesian Optimization with Gaussian Processes for automated configuration tuning of stream processing systems. \citeauthor{Fischer2015} evaluated this method for Apache Storm~\cite{Toshniwal2014} and compared it to a parallel linear (steepest) ascent optimizer, observing comparable performance for both small and large deployments.
\citeauthor{Jamshidi2016} extended this idea by initializing the search with Latin Hypercube Design and showed on Apache Storm benchmarks that this further improves optimization effectiveness over competing methods.
\citet{Bilal2017} proposed a hybrid approach that combines a modified Hill Climbing algorithm with an adapted Latin Hypercube Sampling technique for discrete configuration spaces and a rule‑based component. The method alternates between sampling‑based exploration and local search and experiments with Apache Storm showed improvements over baseline Hill Climbing.

\subsubsection*{Adaptive and Learning‑Based Approaches}

Several approaches for runtime adaptation of stream processing systems are discussed in the literature~\cite{Cardellini2022,Vogel2022}.
For example, \citet{Vysotska2024} propose a \textit{Holistic Adaptive Optimization Technique} that integrates machine learning to continuously monitor system behavior and adapt configuration parameters at runtime, aiming at end‑to‑end performance improvements in heterogeneous streaming architectures rather than tuning single components in isolation.
Along these lines, \citet{Bashtovyi2025} argue that static configuration mechanisms in modern frameworks such as Kafka Streams limit efficiency under variable workloads. They propose an adaptive configuration module that adjusts parameters when performance deviations are detected, combining threshold‑based rules for clear cases with a fine-tuned Large Language Model to select configuration values in ambiguous situations.

\subsubsection*{Discussion}

While adaptive approaches, often based on machine-learning, reduce manual intervention and respond to workload dynamics at runtime, they are less suited to determining a stable configuration at deploy time~\cite{Herodotou2020}.
We instead follow an experiment-driven approach, particularly suited to scenarios without prior configuration knowledge and that require controlled, reproducible executions under well-defined workloads~\cite{Herodotou2020}.
In contrast to the related work mentioned, our method combines the search techniques Latin Hypercube Sampling, Simulated Annealing, and Hill Climbing (see below).
Moreover, we target cloud-native Kubernetes environments and focus on Kafka Streams as a framework specifically suited for stream processing in cloud-native applications, such as event-driven microservices~\cite{Laigner2025,JSS2024}.

\subsection{Employed Search Techniques}

While many search-based optimization techniques exist, our approach builds on the following three:

\textit{Latin Hypercube Sampling} is a statistical technique for generating representative configuration sets in high‑dimensional parameter spaces. Each parameter range is divided into equal‑probability intervals, and samples are constructed so that, for every parameter, each interval is used exactly once across all samples, yielding a space‑filling set of configurations with relatively few samples~\cite{lhsConstraints}.

\textit{Simulated Annealing} is a metaheuristic that explores the configuration space by iteratively proposing small modifications to a current configuration and probabilistically deciding whether to accept them. The decreasing \textit{temperature} controls this probability: high temperatures initially allow frequent acceptance of worse configurations for exploration, while lower temperatures reduce such acceptances and concentrate the search on improvements~\cite{Aarts2005}.

\textit{Hill Climbing} is a similar technique that repeatedly proposes neighboring configurations, but accepts only those that improve the objective~\cite{Bilal2017, Selman2006}.

\section{Performance Optimization Approach}\label{sec:approach}

We propose an approach that automatically explores configuration spaces of stream processing applications to identify settings that optimize performance metrics such as throughput. As input, it takes (i) a stream processing application or benchmark and (ii) a specification of configuration parameters for the application or an underlying streaming framework.

The optimization proceeds iteratively: at each step, the approach proposes a new configuration, runs the application or benchmark under it, and records performance metrics. To balance exploration and exploitation, the search is structured into three phases that employ complementary optimization techniques: Latin Hypercube Sampling for initial space-filling exploration, Simulated Annealing for guided stochastic search, and Hill Climbing for local refinement around promising configurations. The entire workflow is integrated with the Theodolite benchmarking framework~\cite{EMSE2022}, which orchestrates the automated execution of experiments and collection of measurement data.

\subsection{Phase 1: Latin Hypercube Sampling}

The goal of the initial phase is to identify a diverse set of configurations that achieve performance comparable to or better than the default configuration. To this end, the approach first generates a set of candidate configurations using a maximin Latin Hypercube Sampling (LHS) implementation, based on the pre-defined parameter specification. Each sample is an $n$-dimensional point, where each dimension corresponds to one configuration parameter.
The LHS procedure produces normalized values in the range $[0, 1]$ for each dimension. These normalized values are then mapped to the actual parameter ranges using the distribution specified in the parameter definition (e.g., linear or logarithmic scaling). Each resulting configuration is executed while its performance is measured. Finally, the measured performance of each sampled configuration is compared to that of the default configuration, and the best-performing configurations are selected as starting points for the subsequent optimization phase.

\subsection{Phase 2: Simulated Annealing}

The second phase applies Simulated Annealing to explore the neighborhood of the most promising configurations from Phase~1. In each iteration, the algorithm modifies a configurable subset of parameters by a small amount and evaluates the resulting configuration. If the new configuration yields higher performance, it becomes the new current solution. Otherwise, it is accepted with a probability that depends on the performance difference and a \textit{temperature} parameter, which is sampled from a distribution. This process is repeated until a configurable number of iterations is reached. A higher initial temperature increases the likelihood of accepting worse configurations, thereby improving exploration of the configuration space before gradually focusing on better regions.

\subsection{Phase 3: Hill Climbing}

After Phase~2, the best-performing configuration obtained from Simulated Annealing for each initial sample is used as the starting point for the final, exploitation-focused phase. This phase applies a Hill Climbing algorithm that explores the local neighborhood in combination with a smaller number of parameters to be changed in each iteration. In contrast to Simulated Annealing, only configurations that improve performance are accepted, driving the search toward a local optimum around the best-known configuration.

\subsection{Cloud-native Integration}

\begin{figure}
    \centering
    \includegraphics[width=\linewidth]{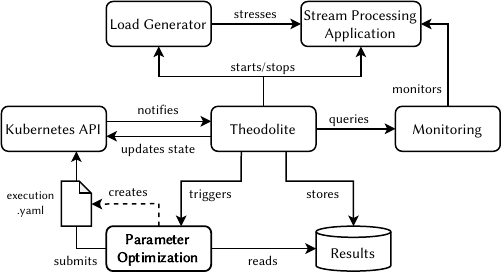}
    \caption{Integration of our optimization approach with Theodolite' cloud-native architecture~\cite{EMSE2022}.}
    \label{fig:architecture}
\end{figure}

The proposed approach is implemented as an extension\footnote{\url{https://github.com/DavidChenY/theodolite-paropt}} to the cloud-native Theodolite benchmarking framework to automate the deployment and execution of experiments on Kubernetes, as shown in \cref{fig:architecture}. For each configuration to be evaluated, our optimization workflow generates a benchmark execution manifest~\cite{EMSE2022} in Theodolite's custom resource format and submits it to the Kubernetes cluster. The Theodolite operator then picks up these manifests and orchestrates the corresponding benchmark runs. To reduce resource consumption, cloud costs, and overall experiment time, we additionally integrated an early-stopping mechanism that detects clearly underperforming configurations based on configurable metrics and terminates their execution before completion.

\section{Experimental Pilot Evaluation}\label{sec:evaluation}

We conduct a pilot evaluation of our optimization approach by applying it to a Kafka Streams benchmark in a controlled yet limited experimental cloud setting. While the results are not intended as a comprehensive or fully generalizable study, they provide initial insights into how the three optimization phases interact and what performance improvements they can achieve for a Kafka Streams application.

\subsection{Methodology and Setup}

The evaluation was conducted on a dedicated Kubernetes cluster provisioned via the CloudLab testbed~\cite{Duplyakin2019}, consisting of three \textit{m510} instances. This setup reduces variability between experiments compared to public cloud environments~\cite{FSE2025}. As workload, we used the ShuffleBench~\cite{ICPE2024} benchmark with its Kafka Streams~\cite{Sax2018,Wang2021} implementation. We deploy three Kafka Streams instances, each limited to one CPU core and 16~GiB of memory in Kubernetes. The benchmark configuration follows our previous work~\cite{ICPE2024,FSE2025}, but uses 10~million consumers and 128~byte records to shift load from the Kafka messaging layer to the Kafka Streams framework.
A configuration is tested in an 8-minute experiment (including 3 minutes of warm-up), and performance is optimized based on the average throughput over this period.

For selecting Kafka Streams configuration parameters, we follow a vendor white paper~\cite{Yewa2020} and the official Kafka Streams documentation. In total, we identified nine parameters (including one RocksDB parameter, three Kafka consumer parameters, and two Kafka producer parameters), listed in \cref{tab:config-results}. For seven of these nine parameters, we apply exponential scaling when mapping from a normalized configuration space to the actual parameter ranges.

\subsection{Evaluation of Optimization Phases}

We analyze the contribution of each optimization phase to the overall search process and the achieved performance improvements.\footnote{The raw data of experiments are provided as supplemental material ~\cite{ReplicationPackage}.}

\begin{table*}
    \centering
    \caption{Best performing configurations after each phase, sorted by throughput. Configuration $c_{x,y,z}$ denotes the $x$-th sample of the Latin Hypercube Sampling phase, the $y$-th iteration of Simulated Annealing, and the $z$-th iteration of Hill Climbing.}
    \label{tab:config-results}
    \footnotesize%
    \newcommand*\rot{\rotatebox{90}}
    \newcommand*{\conf}[1]{\rot{\raggedright\parbox{5.8em}{\raggedright #1}}}
    \newcommand{\dt}{.\allowbreak}
    \newcommand*{\barplot}[1]{\raisebox{\dimexpr-\height+\ht\strutbox}[0pt][0pt]{\includegraphics{#1}}}
    \newcommand*{\barplotlast}[1]{\raisebox{\dimexpr-\height+\ht\strutbox}{\includegraphics{#1}}}
    \begin{tabularx}{\textwidth}{@{}lRRRRRRRRRlr@{}}
        \toprule
        & \multicolumn{9}{l}{Kafka Streams configuration parameter}& \multicolumn{2}{l}{\hspace{3.5pt}Results} \\
        \cmidrule{2-10}
        \cmidrule(l{3.5pt}){11-12}
        Identifier & \conf{cache\dt max\dt bytes\dt buffering} & \conf{buffered\dt records\dt per\dt partition} &  \conf{consumer\dt fetch\dt min\dt bytes} & \conf{commit\dt interval\dt ms} & \conf{producer\dt linger\dt ms} & \conf{consumer\dt{max\dt partition\dt}fetch\dt bytes} & \conf{producer\dt batch\dt size} & \conf{consumer\dt max\dt poll\dt records} & \conf{write\dt buffer\dt size (RocksDB)} & \hspace{3.5pt}Throughput (records/s) & \conf{ $+$/$-$ to previ- ous phase} \\
        \midrule

        $c_{29}$ & 179.4\,KB & 2.9\,M & 2.1\,KB & 5\,258 & 55 & 2.3\,MB & 73\,B & 1.3\,G & 652.1\,MB & \barplot{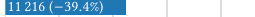} & \\
        $c_{12}$ & 2.1\,KB & 1.2\,G & 127.8\,KB & 4\,359 & 121 & 227.3\,KB & 148\,B & 165.5\,M & 50.6\,MB & \barplot{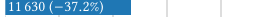} & \\
        $c_{20}$ & 766.7\,KB & 1\,835 & 503\,B & 580 & 166 & 220.0\,MB & 335\,B & 24 & 37.7\,MB & \barplot{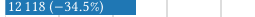} & \\
        $c_{18}$ & 12\,B & 24.0\,M & 113.5\,MB & 2\,326 & 198 & 459.6\,KB & 4.6\,KB & 2 & 255.4\,MB & \barplot{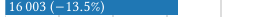} & \\
        $c_{24}$ & 9.2\,MB & 9\,223 & 3.2\,KB & 856 & 48 & 7.8\,MB & 52.4\,KB & 517 & 1.3\,GB & \barplot{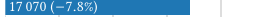} & \\
        $c_{28}$ & 519\,B & 485.1\,K & 30.4\,MB & 5\,137 & 67 & 746.2\,KB & 106.3\,KB & 271 & 1.4\,GB & \barplot{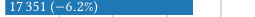} & \\
        $c_{19}$ & 150\,B & 95.8\,K & 16.4\,KB & 3\,475 & 102 & 6.4\,KB & 30.0\,KB & 95.1\,M & 80.2\,MB & \barplot{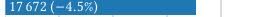} & \\
        $c_\text{default}$ & 10.5\,MB & 1\,000 & 1 & 5\,000 & 0 & 1\,MB &  16\, KB & 500 & 4\,MB & \barplot{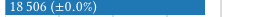}& \\  
        $c_{10}$ & 262.5\,MB & 44.8\,M & 8.6\,MB & 6\,621 & 97 & 932.6\,KB & 4.0\,KB & 57.3\,K & 69.0\,MB & \barplot{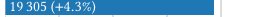} & \\
        $c_{23}$ & 377.5\,KB & 3\,089 & 1.1\,MB & 1\,400 & 189 & 143.6\,KB & 709.4\,KB & 219.2\,K & 1.9\,GB & \barplot{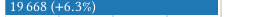} & \\
        $c_{11}$ & 63.1\,KB & 4\,485 & 874\,B & 1\,646 & 145 & 25.0\,MB & 202.5\,KB & 2\,621 & 101.4\,MB & \barplot{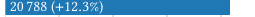} & \\
        $c_{1}$ & 72.6\,MB & 69.3\,M & 143.3\,MB & 4\,195 & 209 & 4.8\,MB & 122.1\,KB & 10.2\,K & 1.1\,GB & \barplot{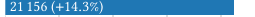} & \\
        $c_{27}$ & 2\,B & 4 & 5.3\,MB & 3\,716 & 216 & 510.1\,MB & 766.7\,KB & 2.4\,M & 183.0\,MB & \barplot{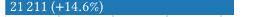} & \\
        
        $c_{10,18}$ & 2.3\,MB & 10.4\,M & 8.6\,MB & 6\,788 & 172 & 2.0\,MB & 4.5\,KB & 23.8\,K & 66.6\,MB & \barplot{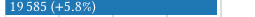} & $+1.5\%$ \\
        $c_{19,17}$ & 24\,B & 5\,937 & 16.4\,KB & 3\,968 & 338 & 6.4\,KB & 39.1\,KB & 136.2\,M & 110.1\,MB & \barplot{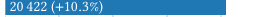} & $+15.6\%$ \\
        $c_{23,14}$ & 54.8\,KB & 24.6\,K & 1.1\,MB & 2\,013 & 381 & 239.4\,KB & 425.8\,KB & 8.2\,M & 1.7\,GB & \barplot{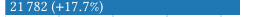} & $+10.7\%$ \\
        $c_{27,20}$ & 2\,B & 32 & 5.3\,MB & 2\,598 & 310 & 512.0\,MB & 1.0\,MB & 41.8\,K & 167.5\,MB & \barplot{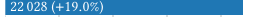} & $+3.8\%$ \\
        $c_{11,7}$ & 13.0\,KB & 2\,440 & 874\,B & 1\,176 & 312 & 33.6\,MB & 202.5\,KB & 25 & 112.7\,MB & \barplot{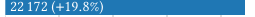} & $+6.6\%$ \\
        $c_{1,13}$ & 71.2\,MB & 28.7\,M & 143.3\,MB & 3\,124 & 442 & 22.7\,MB & 122.6\,KB & 14.2\,K & 1.1\,GB & \barplot{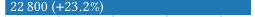} & $+7.8\%$ \\
        
        $c_{19,17,16}$ & 38\,B & 5\,937 & 16.4\,KB & 3\,968 & 451 & 6.4\,KB & 39.1\,KB & 136.2\,M & 147.0\,MB & \barplot{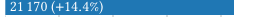} & $+3.7\%$ \\
        $c_{23,10,4}$ & 113.6\,KB & 35.4\,K & 1.1\,MB & 1\,436 & 381 & 239.4\,KB & 547.0\,KB & 8.2\,M & 1.6\,GB & \barplot{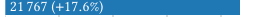} & $+1.2\%$ \\
        $c_{27,20,15}$ & 2\,B & 32 & 5.3\,MB & 2\,598 & 310 & 512.0\,MB & 1.0\,MB & 102.0\,K & 167.5\,MB & \barplot{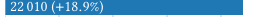} & $-0.1\%$ \\
        $c_{11,7,11}$ & 13.0\,KB & 2\,440 & 874\,B & 495 & 312 & 33.6\,MB & 202.5\,KB & 25 & 112.7\,MB & \barplot{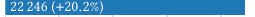} & $+0.3\%$\\
        $c_{1,13,8}$ & 71.2\,MB & 28.7\,M & 143.3\,MB & 3\,051 & 442 & 22.7\,MB & 122.6\,KB & 14.2\,K & 1.1\,GB & \barplotlast{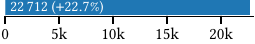} & $-0.4\%$ \\
        \bottomrule
    \end{tabularx}
\end{table*}

\subsubsection{Latin Hypercube Sampling}

In the first phase, we generate an initial set of configurations using maximin Latin Hypercube Sampling with five iterations, resulting in 30 samples. As a baseline, we additionally include the default configuration. Each configuration is executed three times. To filter out faulty, unstable, or clearly low-performing configurations, we employ early termination: executions are stopped if the throughput remains below 30\% of the baseline for 90~seconds or below 50\% of the baseline for 5~minutes.

Out of the 30 generated configurations, 18 are terminated early. Among the 12 configurations that completed, only five yield higher throughput than the baseline, as shown in \cref{tab:config-results}. The worst configuration decreases throughput by about 39\%, whereas the best outperforms the baseline by roughly 15\%. A deeper inspection indicates that particularly low values for \textit{consumer.max.partition.fetch.bytes} frequently lead to early terminations.

\begin{figure}
    \centering
    \includegraphics[width=\linewidth]{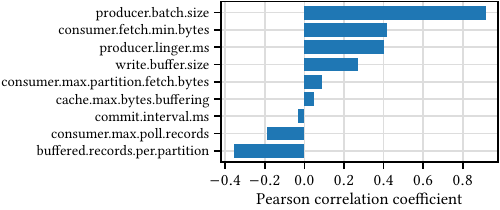}
    \caption{Correlation of parameters with throughput.}
    \label{fig:lhs-correlation}
\end{figure}

For the 12 fully executed configurations, we conduct a correlation analysis between configuration parameters and measured throughput, as shown in \cref{fig:lhs-correlation}. We observe a very strong positive correlation for \textit{producer.batch.size} and moderately strong positive correlations for \textit{producer.linger.ms} and \textit{consumer.fetch.min.bytes}. In addition, we find a weak negative correlation for \textit{buffered.records.per.partition}.

\subsubsection{Simulated Annealing}

For the Simulated Annealing phase, we continue with the six configurations that performed similarly to or better than the default configuration. For each of these configurations, we run Simulated Annealing for 25 iterations, resulting in 150 evaluated configurations in total. We derive the initial temperature from a specified throughput loss of 2\,500 records/s (about 14\% below the baseline throughput) that we still want to accept with 75\% probability at the start, using the Boltzmann equation, and employ the widely used exponential cooling function with a cooling rate of 0.95.
As this phase is deliberately focused on exploration rather than exploitation, we choose such a high acceptance rate for worse configurations and do not use repetitions.
In every iteration, two parameters are modified by a value in the range of $[-10\%, +10\%]$ on the normalized parameter scale.

\begin{figure}%
	\centering%
    \hspace{0.319447in}%
    \subcaptionbox{$c_{1}$}{%
        \hspace{-0.319447in}%
        \includegraphics{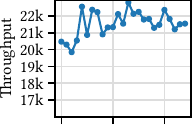}%
    }\hfill%
    \subcaptionbox{$c_{10}$}{%
        \includegraphics{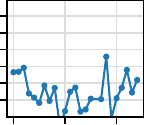}%
    }\hfill%
    \subcaptionbox{$c_{11}$}{%
        \includegraphics{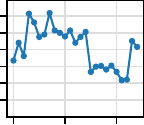}%
    }
    
    \vspace{0.5em}

    \hspace{0.319447in}%
    \subcaptionbox{$c_{19}$}{%
        \hspace{-0.319447in}%
        \includegraphics{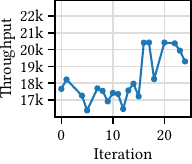}%
    }\hfill%
    \subcaptionbox{$c_{23}$}{%
        \includegraphics{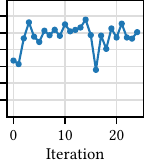}%
    }\hfill%
    \subcaptionbox{$c_{27}$}{%
        \includegraphics{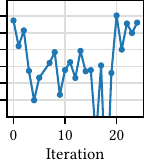}%
    }%
	\caption{Evolution of throughput (records/s) over Simulated Annealing iterations for different starting configurations.}
    \label{fig:sa-iterations}
\end{figure}

Across all Simulated Annealing runs, only seven executions are terminated early, five of which stem from the lowest-performing configuration ($c_{19}$) in Phase~1. \Cref{fig:sa-iterations} shows how the observed throughput evolves over the iterations. The high acceptance rate (together with the inherent performance variability) is visible in frequent decreases in throughput. Nevertheless, throughput increases from the first to the last iteration for four out of the six starting configurations. Overall, the highest throughput is achieved in the 13th iteration for configuration~$c_{1}$, with a gain of approximately 23\% compared to the baseline configuration and about 8\% compared to its own starting point in this phase (see again \cref{tab:config-results}).

\subsubsection{Hill Climbing}

In the third phase, we shift the focus from exploration to exploitation and use Hill Climbing to fine-tune configurations around the best results from Simulated Annealing.
As none of the iterations on $c_{10}$ reached a throughput of 20\,000 records/s, they are excluded from further fine-tuning.
To reflect the focus on local refinement, we reduce the number of iterations to 17. In each iteration, a single parameter is modified by a value in the range of $[-10\%, +10\%]$ on the normalized parameter scale.

\begin{figure}%
	\centering%
    \hspace{0.319447in}%
    \subcaptionbox{$c_{1,13}$}{%
        \hspace{-0.319447in}%
        \includegraphics{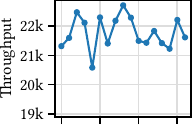}%
    }\hfill%
    \subcaptionbox{$c_{11,7}$}{%
        \includegraphics{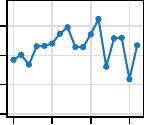}%
    }\hfill%
    \subcaptionbox{$c_{19,17}$}{%
        \includegraphics{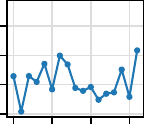}%
    }
    
    \vspace{0.5em}

    \hspace{0.319447in}%
    \subcaptionbox{$c_{23,10}$}{%
        \hspace{-0.319447in}%
        \includegraphics{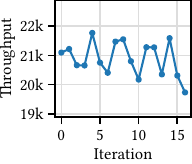}%
    }\hfill%
    \subcaptionbox{$c_{27,20}$}{%
        \includegraphics{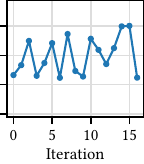}%
    }\hfill%
    \phantom{\includegraphics{images/hc-iterations_3.pdf}}%
	\caption{Evolution of throughput (records/s) over Hill Climbing iterations for different starting configurations.}%
    \label{fig:hc-iterations}%
\end{figure}

\Cref{fig:hc-iterations} summarizes the results of the Hill Climbing phase. The four best configurations all stem from $c_{1,13}$, which was already the best-performing configuration in the Simulated Annealing phase. However, across all starting points, we observe no substantial improvements (see again \cref{tab:config-results}). Although Hill Climbing is designed to accept only better configurations, the throughput still fluctuates slightly up and down over the iterations, indicating that general result variability dominates.
Re-executing both the starting configuration and the best Hill Climbing configuration with three repetitions each yields differences of only $-4.2\%$ to $+2.9\%$, confirming that the apparent gains lie within the natural variability of the benchmark.

\subsection{Discussion}

Our pilot evaluation shows that the proposed approach can substantially improve the throughput of a Kafka Streams application, achieving a gain of about 23\% over the default configuration. In particular, the combination of Latin Hypercube Sampling and Simulated Annealing proved effective in identifying and refining promising regions of the configuration space, whereas the final fine-tuning step with Hill Climbing did not yield noticeable additional improvements beyond the natural variability of the benchmark.

A critical factor for practical applicability is the execution time of experiments. Here, the early termination mechanism is highly beneficial: in our first phase's experiments, around 60\% of configurations were stopped early, reducing overall execution time by more than 50\%.
We recently quantified performance variability for the same benchmark in public clouds~\cite{FSE2025}, which opens up opportunities to incorporate variability information into the search process.

\begin{figure}
    \centering
    \includegraphics{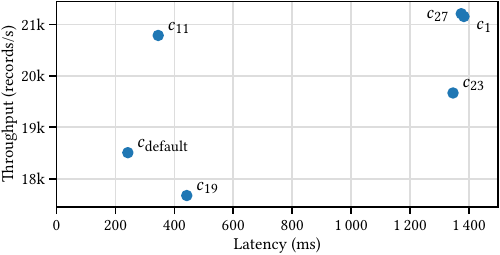}
    \caption{Latency and throughput of selected configurations.}
    \label{fig:lhs-latency-throughput}
\end{figure}

Our presented pilot evaluation focuses on throughput as the primary objective. In realistic scenarios, however, other properties such as latency and fault tolerance are also relevant. While fault-tolerance experiments are comparatively complex~\cite{DEBS2024,Tahir2024}, we conducted an exploratory latency analysis for selected configurations as shown in \cref{fig:lhs-latency-throughput}. Although all configurations that improve throughput over the default configuration also increase latency, we do not observe a clear correlation between absolute throughput and latency, suggesting opportunities for multi-objective optimization that explicitly balances these trade-offs in future work.

\section{Conclusions and Outlook}\label{sec:conclusions}

This paper presents our automated, experiment-driven approach for configuration optimization of stream processing systems, with a special focus on cloud-native deployments such as Kubernetes. In an experimental evaluation, our approach was able to improve throughput of a Kafka Streams benchmark by up to 23\% over the default configuration.
Most of these gains were achieved by the combination of Latin Hypercube Sampling and Simulated Annealing, together with early termination of underperforming or faulty configurations. Additional fine-tuning via Hill Climbing yielded only marginal improvements.
For future work, we plan more extensive evaluations in public cloud environments, extend the approach to multi-objective optimization (e.g., throughput and latency), and apply it to other stream processing frameworks.

\begin{acks}
We would like to thank the Johannes Kepler University Linz and Dynatrace for co-funding this research.
\end{acks}
\balance

\bibliographystyle{ACM-Reference-Format}
\bibliography{references.bib}

\end{document}